\documentclass[aps,pra,twocolumn,groupedaddress,showpacs]{revtex4}

\usepackage{graphicx}
\usepackage{amsmath,amssymb}
\usepackage{amsfonts}
\usepackage{color}

\begin{document}

\title{Robust Unconditionally Secure Quantum Key Distribution\\
with Two Nonorthogonal and Uninformative States}

\author{Marco~Lucamarini$^{1}$}

\author{Giovanni~Di~Giuseppe$^{2}$}

\author{Kiyoshi~Tamaki$^{3,4}$}

\affiliation{$^{1}$CNISM UdR University of Camerino, Via Madonna
delle Carceri 9, 62032 Camerino (MC), Italy.\\
$^{2}$Physics Department, University of Camerino, Via Madonna
delle Carceri 9, 62032 Camerino (MC), Italy.\\
$^{3}$NTT Basic Research Laboratories, NTT Corporation 3-1,
Kanagawa, 243-0198, Japan\\
$^{4}$CREST, JST Agency, 4-1-8 Honcho, Kawaguchi, Saitama,
332-0012, Japan.}

\date{\today}

\begin{abstract}
We introduce a novel form of decoy-state technique to make the
single-photon Bennett 1992 protocol robust against losses and
noise of a communication channel. Two uninformative states are
prepared by the transmitter in order to prevent the unambiguous
state discrimination attack and improve the phase-error rate
estimation. The presented method does not require strong reference
pulses, additional electronics or extra detectors for its
implementation.
\end{abstract}

\pacs{03.67.Dd, 03.67.Hk}
\maketitle

\section{INTRODUCTION}

Quantum Key Distribution (QKD) is a way to distribute secret keys
between two distant parties with provable
security~\cite{GRT+02,May96,LC99,SP00}. Despite the general
principles of QKD are now well known, there is no definite answer
yet about its effective use in the real world, being it dependent
on practical figures of merit like the transmission rate, the
working distance and even the production cost of a specific
implementation. A useful criterion in such intricate situations is
that of the simplest choice. In case of cryptography this is
particularly desirable for it allows general and reliable security
proofs together with efficient and low-cost practical
realizations.

The simplest QKD protocol was conceived by C. H. Bennett in 1992
and named after him ``B92''~\cite{Ben92}. It is based on only two
nonorthogonal states associated with the two values of the logical
bit to-be-secretly-transmitted.
%
%
Despite its simplicity, the B92 is considered as a quite
impractical protocol, mainly for its low tolerance to the losses
and noise of a communication channel~\cite{UpQC}. The high
dependance on channel losses can be ascribed to the so-called
Unambiguous State Discrimination (USD) attack~\cite{DLH06}, which
represents the principal threat against the B92, and severely
limits its performances. On the other side, the high dependance on
the channel noise can be imputed to the lack of a direct
phase-error estimation, which entails a more than necessary
privacy amplification as soon as the noise sensibly increases.

So far, no solution has been devised to improve the phase-error
estimation of B92, and only one solution is available to contrast
the USD attack. It was originally proposed in~\cite{Ben92}, and
recently proved unconditional secure in~\cite{Koa04}
and~\cite{TLK+06}. It consists of a strong reference pulse
accompanying the signal pulse and phase-related to it. The
presence of the reference pulse prevents an eavesdropper (Eve)
from selectively stopping the signals according to the results of
an USD measurement. By adopting the technology reported
in~\cite{GYS04}, this solution would allow a secure QKD over
distances of about $124$~Km~\cite{TLK+06}. However the high
intensity of the reference pulse is expected to cause in practice
noise due to the scattering inside an optical fiber~\cite{SZT05},
while if a weaker reference pulse is used~\cite{Tam08}, the
maximum distance drops to about $87$~Km. Furthermore the solution
of a strong reference pulse cannot be applied if the B92 is
realized with a single-photon source or with SPDC (Spontaneous
Parametric Down Conversion), because the beam intensities cannot
be modulated to the necessary extent. On the contrary, a solution
functioning at the single-photon level could be at once
transferred to a realistic scenario by virtue of the decoy-state
technique~\cite{Hwa03,LMC05,Wan05}, which allows the estimation of
statistical quantities related to the single-photon pulses only.

\begin{figure}[h!]
\includegraphics[width=9.0cm]{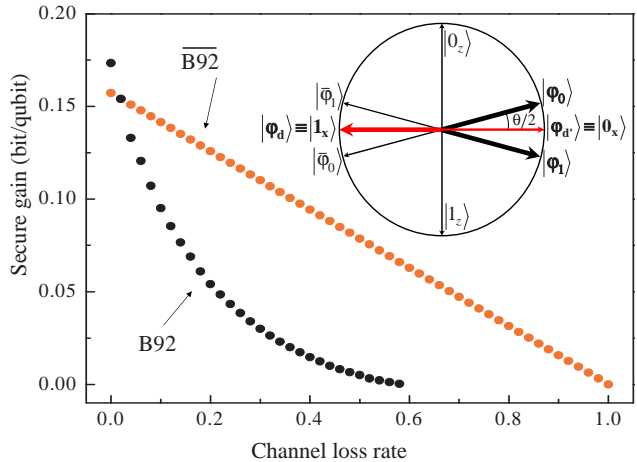}\newline
\caption{(color online) Plots of the secure gain $G$ as a function
of the channel loss rate $L$, for the standard B92 (black circles)
and for the $\overline{\textrm{B92}}$ (red circles). The
$\overline{\textrm{B92}}$ scales linearly with $L$, while the B92
scales nearly quadratically, and ceases to provide a positive gain
at about $L=0.6$. The curves are drawn assuming a depolarizing
channel with $p=0.01$ (see the text). On the top-right, the states
of traditional B92 (black arrows) and the extra uninformative
states (red arrows) used in the two modifications of the B92 (see
the text).} \label{fig1}
\end{figure}

In this paper we propose a novel solution to make the
single-photon B92 robust against the losses and the noise of a
communication channel. First, the transmitter prepares, besides
the two conventional signal states, two additional uninformative
states. This modification alone is very feasible and allows to
detect the USD attack. The protocol containing this first
modification will be called $\overline{\textrm{B92}}$ henceforth.
In contrast to the state-of-the-art, the gain~\cite{Note2} of the
single-photon $\overline{\textrm{B92}}$ depends only linearly on
the loss rate of a communication channel rather than nearly
quadratically. This result is anticipated in Fig.~\ref{fig1}.
%
%
As a second step, the receiver's box is slightly modified so to
directly measure the phase-error of the B92, whose estimation is
known to be quite poor at large angles between the two signal
states. The protocol containing both the first and the second
modifications will be called $\overline{\overline{\textrm{B92}}}$.
This further variant provides a higher tolerance to the channel
noise, approaching, when the two signal states are nearly
orthogonal, the one featured by the BB84~\cite{BB84}.

The main achievement of the paper is the formulation of the
$\overline{\textrm{B92}}$. It relies on the observation that the
states of B92, being only two, are always linearly independent.
Since the USD attack is effective only on sets of linearly
independent states~\cite{Che98}, it is quite natural to introduce
one more state to obtain a set of linearly dependent states. This
removes at the root the possibility of an USD attack. Actually the
states added in the new protocol are two rather than one; this
depends on technical reasons related to the unconditional security
proof which will become clear later. The added states are
uninformative and after the quantum transmission are discarded.
Even so the users can measure the loss-rate pertaining to these
states and obtain a signature of Eve's presence. This solution
resembles the decoy-state technique~\cite{Hwa03,LMC05,Wan05} used
to realize a long-distance BB84 in a realistic scenario; so we
will call the extra uninformative states ``decoy states''
henceforth. However it should be noted that the conventional
decoy-state comes from the intensity modulation of a certain
pulse, i.e. from the modulation of a degree of freedom different
from the one in which information is encoded; on the contrary, in
the present case, the decoy-state is encoded in the same degree of
freedom of the signal states, and there is no modulation in other
degrees of freedom different from that. This is a relevant
peculiarity of the $\overline{\textrm{B92}}$ since it entails that
no additional hardware respect to a standard B92 setup is required
to implement the new protocol.

In the next, we focus on the $\overline{\textrm{B92}}$, leaving
the $\overline{\overline{\textrm{B92}}}$ for the conclusion of the
paper. It should be noted however that all the results obtained
for the former protocol, in particular its unconditional security,
hold for the latter protocol as well; so they will be given
without repeating unnecessary security proofs.

Our work is structured as follows: in
Section~\ref{sec:preliminary} we briefly review the security proof
of the standard B92, to intuitively explaining the idea of using
uninformative states to protect against the USD attack. Then, in
Section~\ref{sec:uncsec}, we introduce the
$\overline{\textrm{B92}}$ and provide the proof of its
unconditional security. In Section~\ref{sec:numsim} we show, with
the help of numerical simulations, the independence of the new
protocol from the losses of a communication channel. In
Section~\ref{sec:phase-err}, we detail the
$\overline{\overline{\textrm{B92}}}$, in which the users perform a
direct estimate of the phase-error rate. The concluding remarks
are given in Section~\ref{sec:conclusion}.

\section{PRELIMINARY CONSIDERATIONS}
\label{sec:preliminary}

Let us introduce the notation to explain the B92 protocol. We
write the bit encoded by the transmitter (Alice) as $j=\{0,1\}$
and the corresponding qubit as $\left\vert \varphi
_{j}\right\rangle \equiv \beta \left\vert 0_{x}\right\rangle
+\left( -1\right) ^{j}\alpha \left\vert 1_{x}\right\rangle $,
where $\left\{ \left\vert 0_{x}\right\rangle ,\left\vert
1_{x}\right\rangle \right\} $ are the eigenstates of the $X$
basis, $\beta =\cos \frac{\theta }{2}$, $\alpha =\sin \frac{\theta
}{2}$, $0<\theta <\pi /2$ (see Fig.\ref{fig1}). The bases $X$ and
$Z$ are related by $\left\vert j_{z}\right\rangle =\left[
\left\vert 0_{x}\right\rangle +(-1)^{j}\left\vert
1_{x}\right\rangle \right] /\sqrt{2}$. We can also introduce the
state $ \left\vert \overline{\varphi
}_{j}\right\rangle=\alpha|0_{x}\rangle-(-1)^{j}\beta|1_{x}\rangle$
orthogonal to $\left\vert \varphi_{j}\right\rangle $. It is
possible now to define a general USD measurement, parametrized by
$\gamma$, as $\mathcal{M}_{B92}^{\gamma } \equiv \left\{
F_{0}^{\gamma},F_{1}^{\gamma },F_{inc}^{\gamma},F_{v}\right\} $,
with $F_{0}^{\gamma }=\left( \frac{\gamma ^{2}}{2\beta
^{2}}\left\vert \overline{\varphi }_{1}\right\rangle \left\langle
\overline{\varphi }_{1}\right\vert \right)$, $F_{1}^{\gamma
}=\left( \frac{\gamma ^{2}}{2\beta ^{2}}\left\vert
\overline{\varphi }_{0}\right\rangle \left\langle
\overline{\varphi }_{0}\right\vert \right) $, $F_{inc}^{\gamma
}=1-F_{0}^{\gamma }-F_{1}^{\gamma }$, $F_{v}=\left\vert v
\right\rangle\left\langle v \right\vert$; $\left\vert v
\right\rangle$ is a state that includes both vacuum and
multi-photon pulses, whose occurrences depends on $L$, the total
loss-rate of the communication channel, which is under Eve's
control.
It is maybe useful to point out that the operators
$F_{0}^{\gamma}$, $F_{1}^{\gamma}$ and $F_{inc}^{\gamma}$ live in
the subspace characterized by the single-photon projector $\Pi_s$,
while $F_{v}$ lives in the complementary subspace $1-\Pi_s$.
Note that we are assuming for simplicity a perfect single-photon
source for Alice and ideal detectors for Bob. More precisely Bob's
detectors can discriminate vacuum, single-photon and multi-photon
pulses, and they are modeled like a beam-splitter with
transmission $\eta_{B}$, controlled by Eve, followed by detectors
with unit efficiency. The control by Eve on detectors is taken
into account by including $\eta_{B}$ in the loss-rate $L$. We
remark that the assumption of a perfect single-photon source can
be dropped if the standard decoy-state
technique~\cite{Hwa03,LMC05,Wan05} is brought into the
description. Also, we define for later use $F_{conc}^{\gamma
}=\left(F_{0}^{\gamma }+F_{1}^{\gamma }\right) =
(A_{fil}^{\gamma})^2$, and we include in the term ``vacuum'' both
the vacuum and the multi-photon events. Finally, we observe that
the measurement $\mathcal{M}_{B92}^{\gamma}$ is \textit{optimal}
when $\gamma =1$, while it is \textit{practical} when $\gamma
=\beta $~\cite{Note3}.
While the receiver (Bob), due to his limited technology, can
execute just the practical measurement
$\mathcal{M}_{B92}^{\beta}$, Eve is supposed to be endowed with
superior technology, so she can execute the optimal
$\mathcal{M}_{B92}^{1}$, with even the additional condition $L=0$.
In fact we conservatively assume that Eve completely controls the
channel, so all the losses and noises are caused by her.

%
\indent The unconditional security of a lossless B92 was proved
for the first time in~\cite{TKI03} and that of a lossy B92
in~\cite{TL04}. It is shown there that the Prepare-and-Measure
(PM) B92, experimentally accessible to Alice and Bob, can be
obtained through a reduction argument by another protocol based on
Entanglement Distillation (ED)~\cite{BDS+96}. The ED protocol is
not really implemented by the users; it is rather a theoretical
tool to find the conditions under which the B92 operations can
lead to the distillation of the maximally entangled state
\begin{equation}
\left\vert \Phi^{+}\right\rangle _{AB}=\frac{1}{\sqrt{2}}\left(
\left\vert 0_{x}\right\rangle _{A}\left\vert 0_{x}\right\rangle
_{B}+\left\vert 1_{x}\right\rangle _{A}\left\vert
1_{x}\right\rangle _{B}\right),  \label{Phi}
\end{equation}
starting from the state
\begin{align}
\left\vert \Psi\right\rangle _{AB}  & =\frac{1}{\sqrt{2}}\left(
\left\vert 0_{z}\right\rangle _{A}\left\vert
\varphi_{0}\right\rangle _{B}+\left\vert 1_{z}\right\rangle
_{A}\left\vert \varphi_{1}\right\rangle _{B}\right)
\nonumber\\
& =\beta\left\vert 0_{x}\right\rangle _{A}\left\vert
0_{x}\right\rangle _{B}+\alpha\left\vert 1_{x}\right\rangle
_{A}\left\vert 1_{x}\right\rangle _{B},\label{InitState}
\end{align}
which explicitly contains the two signal qubits.

The reduction from the ED to the PM protocol is obtained by noting
that the measurement $\mathcal{M}_{B92}^{\beta}$ effected by Bob
in the PM protocol is equivalent in the ED protocol to a $Z$-basis
measurement conditional on the successful filtering operations
$Q_{v}\equiv\left\{ F_{v},1-F_{v}\right\}$ and $A_{fil}^{\beta}$.
In particular, when the filters are successful, the Bell
state~\eqref{Phi} is obtained with probability $2 \alpha^2
\beta^2$ by the state~\eqref{InitState}, if the channel is
lossless and noiseless~\cite{BBP+96}. Once the users share the
state (\ref{Phi}) they can obtain a secure key by the subsequent
measurement in the $Z$ basis.

Nonetheless, in real situations, the channel is not lossless and
noiseless, so the filtered states may include $n_{bit}$ bit
errors, represented by the states $\{|0_z\rangle_A
|1_z\rangle_B,|1_z\rangle_A |0_z\rangle_B\}$, and $n_{ph}$ phase
errors, represented by the states $\{|0_x\rangle_A
|1_x\rangle_B,|1_x\rangle_A |0_x\rangle_B\}$.
Alice and Bob must then resort to an ED based on CSS
codes~\cite{CS95,Ste96} in order to correct the errors and distill
the Bell state~\eqref{Phi}.
This procedure needs not to be actually accomplished by the users;
it suffices that they provide reliable upper bounds to the number
of bit errors and phase errors present in their data. This is the
main breakthrough of the security proof given in~\cite{SP00}.

Bit errors can be directly estimated by sacrificing a part of the
data, which are publicly revealed on the classical channel. On the
contrary phase errors can not be directly estimated in B92. This
is due to the fact the none of the operations of this protocol can
be made equivalent to a measurement in the $X$ basis. To cope with
this problem, it is useful to devise a gedanken experiment in
which Alice and Bob perform a measurement of the
state~\eqref{InitState} in the $X$ basis. In this way the gedanken
outcomes can be put in relation with measurable quantities of the
protocol. Let us indicate with $|i_x\rangle_A |j_x\rangle_B$
($i=\{0,1\},j=\{0,1,v\}$) the gedanken outcomes, and with
$n_{i,j}$ the number of their occurrences. Then, by looking at the
second of Eqs.\eqref{InitState}, and by considering Bob's
measurement $\mathcal{M}_{B92}^{\beta}$, we can easily obtain the
following relations, valid in the asymptotic limit of large
$N$~\cite{TKI03,TL04}:
\begin{eqnarray}
\label{nph1} n_{ph} &=& \beta^2 n_{01} + \alpha^2 n_{10}~,\\
\label{n1v} \alpha^2N &=& n_{10} + n_{11} + n_{1v}~.
\end{eqnarray}
Eq.~\eqref{nph1} quantifies, in terms of gedanken quantities, the
number of phase errors; it contains $n_{01}$ and $n_{10}$ which
can be put in relation to measurable quantities through an
argument based on quantum theory~\cite{TKI03}. Eq.~\eqref{n1v} is
a direct consequence of the fact that neither Eve nor the channel
can access Alice's qubit as long as it is in Alice's hands; it
contains the crucial parameter $n_{1v}$, which quantifies the
\textit{effective nonorthogonality} of the signal states in
presence of losses. In fact, consider the case of a lossless
channel, i.e. $n_{1v}=0$. From Eq.\eqref{n1v} we can see that the
quantity $\Delta=n_{10} + n_{11}$ measures the degree of
nonorthogonality of the two signal states, being it equal to
$\alpha^2N$, and being $1-2\alpha^2$ equal to the scalar product
of the two signal states. When losses are taken into account,
$\Delta$ still represents the nonorthogonality of the two states
on a loss-free channel, but this time it is equal to
$\alpha^2N-n_{1v}$, smaller than before. This means that losses
have increased the orthogonality of the two states, making the B92
more prone to USD and other loss-based attacks. So it is important
to include $n_{1v}$ in the phase estimation process, for its
optimization is directly related to the amount of information
leaked to Eve through a loss-based mechanism. By executing the
numerical search of the phase error upper bound, it turns out that
the optimal value for $n_{1v}$ is \textit{zero} in most of the
cases. Below, we show that it suffices to consider the USD attack
to intuitively explain this value.\\
The USD attack is performed by Eve via the optimal measurement
$\mathcal{M}_{B92}^{1}$, executed on the qubits prepared by Alice.
Let us consider the signal states $\{\left\vert
\varphi_{0}\right\rangle,\left\vert \varphi_{1}\right\rangle\}$
and the $X$-basis states $\{\left\vert
0_{x}\right\rangle,\left\vert 1_{x}\right\rangle\}$. A conclusive
outcome is obtained by Eve for these four states with the
following probabilities:
%
%
\begin{eqnarray}
\label{Rates1} P_{\varphi_0}^{USD} &=& P_{\varphi_1}^{USD} = 1-\cos \theta,\\
\label{Rates2} P_{0_x}^{USD} &=& 0,~~~P_{1_x}^{USD} = 1.
\end{eqnarray}
If the result is conclusive, Eve forwards the qubit (in the
correct state) to Bob; otherwise she stops the qubit and creates a
loss in the channel. So, while in case of no attack the four
states are expected to arrive at Bob's detectors with the same
probability, in case of USD attack, the users would detect $ N_s
(1-\cos \theta)$ signal states, $N_d$ states $\left\vert
1_{x}\right\rangle$ and $0$ states $\left\vert
0_{x}\right\rangle$, where $N_s$ and $N_d$ are respectively the
total number of signal and decoy states prepared by Alice. Note
that this implies in case of USD attack the following setting:
\begin{equation}\label{WCass}
   n_{1v}=0.
\end{equation}
In standard B92 the users can not measure the loss-rates of
Eq.\eqref{Rates2}, so they conservatively assume that Eve executed
an USD attack even if she actually did not, whence
Eq.\eqref{WCass}. But the loss-rates of Eq.\eqref{Rates2} become
measurable in $\overline{\textrm{B92}}$, because of the presence
of two uninformative states which are chosen on purpose equal to
$\left\vert 0_{x}\right\rangle$ and $\left\vert
1_{x}\right\rangle$. This is the main feature that makes the new
protocol independent of losses.


\section{PROTOCOL AND UNCONDITIONAL SECURITY}
\label{sec:uncsec}

In the $\overline{\textrm{B92}}$, Alice prepares the two signal
states $\left\{ \left\vert \varphi _{0}\right\rangle ,\left\vert
\varphi _{1}\right\rangle \right\}$ plus two additional
uninformative decoy states $\{\left\vert \varphi
_{d}\right\rangle,\left\vert \varphi _{d'}\right\rangle\} $, which
are chosen respectively equal to the states $\{\left\vert
1_{x}\right\rangle,\left\vert 0_{x}\right\rangle\} $.
Here we consider two decoy states only, but the use of three or
more decoy states could be useful to adverse other kinds of
USD-based attacks, like those described in~\cite{Koa05,LCD+07}.
Notice that the addition of two decoy states makes the overall
states prepared by Alice linearly dependent; hence is impossible
for Eve to unambiguously discriminate them. More importantly, the
presence of two decoy states, prepared with suitable
probabilities, makes the signal and decoy density matrices
identical. This is fundamental for the unconditional security
proof of the protocol, because it prevents Eve from behaving
differently with signals and decoys, and legitimates the use of
random sampling arguments between the two classes of pulses
prepared by Alice. Let us write explicitly the density matrix of
the signal states:
\begin{eqnarray}\label{DecStProb}
\nonumber
\rho&=&(|\varphi_{0}\rangle\langle\varphi_{0}|+|\varphi_{1}
          \rangle\langle\varphi_{1}|)/2 \\
    &=&\beta^2|0_x\rangle\langle0_x|+\alpha^2|1_x\rangle\langle1_x|.
\end{eqnarray}
The above equation explains why we chose the decoy states equal to
the $X$-basis states $\{\left\vert 1_{x}\right\rangle,\left\vert
0_{x}\right\rangle\}$. Moreover, the requirement of equal density
matrices for signal and decoy states fixes the preparation
probability of the states $\left\vert 1_{x}\right\rangle$ (i.e.
$|\varphi_{d}\rangle$) and $\left\vert 0_{x}\right\rangle$ (i.e.
$|\varphi_{d'}\rangle$) respectively to $\alpha^2$ and $\beta^2$.
After the whole quantum transmission the decoy instances are
discarded. Even so, they allow the direct estimate of the crucial
quantity $n_{1v}$ previously introduced.

For the security proof of $\overline{\textrm{B92}}$ we adopt the
same argument as in~\cite{TL04}: we introduce below the PM
$\overline{\textrm{B92}}$ and show that it can be obtained by an
ED protocol through a reduction argument. Then we show the
unconditional security of the ED protocol by applying it to a
Shor-Preskill security proof~\cite{SP00}.

\bigskip

\textit{PM $\overline{\textrm{B92}}$} -- (i) Alice randomly and
uniformly prepares $2N$ signal qubits in the state $\left\vert
\varphi _{0}\right\rangle $ or $\left\vert \varphi
_{1}\right\rangle $, $\alpha ^{2}N$ decoy qubits in the state
$\left\vert \varphi_{d}\right\rangle $ and $\beta^{2}N$ decoy
qubits in the state $\left\vert \varphi_{d'}\right\rangle $. --
(ii) Bob executes the measurement $\mathcal{M}_{B92}^{\beta }$: in
case he obtains $\left\vert v\right\rangle$ he labels the outcome
as `vacuum'; in case he obtains $\left\vert
\varphi_{0}\right\rangle $ or $\left\vert
\varphi_{1}\right\rangle$ he labels the outcomes as
`inconclusive'; in case he obtains $\left\vert
\overline{\varphi}_{0}\right\rangle $ ($\left\vert
\overline{\varphi}_{1}\right\rangle $) he labels the outcome as
`conclusive' and decodes as `$1$' (`$0$'). -- (iii) After the
quantum transmission, the users randomly permute their bits. Then
Bob tells Alice the positions of vacuum, inconclusive and
conclusive counts. Alice calculates the number $n_{kv}$ of joint
occurrences
$\{\left\vert\overline{\varphi}_{k}\right\rangle,\left\vert v
\right\rangle\}$ ($k=\{0,1,d,d'\}$) between her preparation and
Bob's measurement, the number $n_v=\sum_k n_{kv}$ of total vacuum
counts, and the number $n_{conc}$ of conclusive counts. She
announces the estimated quantities to Bob together with the
positions of decoy bits. -- (iv) The data corresponding to decoy
or inconclusive outcomes are removed by the users. -- (v) The
first half of the remaining bits (check bits) are used for
estimating the number of errors $n_{err}$ in which Alice prepared
$\left\vert \varphi_{0}\right\rangle $ ($\left\vert \varphi
_{1}\right\rangle $) and Bob decoded as `$1$' (`$0$'). -- (vi)
From $n_v$, $n_{err}$, $n_{conc}$, and $n_{dv}$ the users estimate
the number of bit errors $n_{bit}$, and an upper bound on the
number of phase errors $\overline{n}_{ph}$ in the second half of
the remaining bits (data bits). -- (vii) The users perform error
correction and privacy amplification on the data bits according to
the values of $n_{bit}$ and $\overline{n}_{ph}$ respectively, thus
obtaining an $n_{key}$-bit shared secret key.

\bigskip

In the following we provide the ED protocol that we show to be
unconditionally secure, and that will eventually reduce to the PM
$\overline{\textrm{B92}}$.

\bigskip

\textit{ED $\overline{\textrm{B92}}$} -- (i) Alice prepares $3N$
copies of the bipartite state given in Eq.\eqref{InitState} and
sends the $3N$ systems $B$ to Bob over the quantum channel. --
(ii) Alice and Bob randomly permute by public discussion the
positions of all their pairs. -- (iii) Bob performs the QND
measurement described by $Q_{v}=\left\{F_{v},1-F_{v}\right\} $,
and publicly announces the outcomes~\cite{Note6}; let $n_v$ be the
number of outcomes associated with $F_{v}$. -- (iv) For the first
$N$ pairs (decoy pairs), Alice measures system $A$ in the $X$
basis, and publicly announces the positions of the decoy pairs.
Since neither Eve, Bob nor the channel can touch Alice's qubits,
we can infer from Eq.\eqref{InitState} that she will obtain
$\left\vert 0_{x}\right\rangle$ with probability $\beta ^{2}$ and
$\left\vert 1_{x}\right\rangle $ with probability $\alpha^{2}$.
Alice counts the number $n_{1v}$ of joint occurrences
$\{\left\vert 1_{x}\right\rangle $, $F_{v}\}$ in the outcomes.
Then the users discard these results.
%
%
-- (v) For half of the remaining pairs (check pairs), Alice
measures system $A$ in the $Z$ basis, and Bob performs the
measurement $\mathcal{M}_{B92}^{\beta }$ on his system. By public
discussion, they determine the number $n_{err}$ of errors in which
Alice found `$0$' (`$1$') and Bob's outcome was `$1$' (`$0$'). --
(vi) For the other half of the remaining pairs (data pairs), Bob
performs the filtering $A^{\beta}_{fil}$ on each of his qubits,
and announces the positions and the total number $n_{fil}$ of the
qubits that have passed the filtering. -- (vii) From $n_v$,
$n_{err}$, $n_{fil} $ and $n_{1v}$ the users estimate an upper
bound for the number of bit errors $n_{bit}$ and phase errors
$n_{ph}$, in the $n_{fil}$ pairs. -- (viii) They run an ED
protocol that can produce $n_{key}$ nearly perfect EPR pairs if
the estimation is correct. -- (ix) Alice and Bob each measures the
EPR pairs in $Z$ basis to obtain an $n_{key}$-bit shared secret
key.

\bigskip

The unconditional security of our protocol follows the proof given
in~\cite{TL04} after minor modifications. Actually, all the
operations and measurements in our entanglement-based protocol,
with exception of Step (iv), are purposely chosen equal
to~\cite{TL04} to maximally exploit the results obtained there.

The first step is to show the equivalence of the two protocols
given above. For that, it suffices to note that Eve can not
distinguish the preparation of the decoy states $\{\left\vert
\varphi_{d}\right\rangle, \left\vert \varphi_{d'}\right\rangle \}
$ in the PM protocol from that of $\{\left\vert1_{x}\right\rangle,
\left\vert 0_{x}\right\rangle \} $ in the ED protocol, effected
through a $X$ basis measurement, since the resulting states are
the same, they are prepared with the same probabilities and the
time at which Alice performs the $X$ basis measurement can not
have an influence on the results. This also implies that the
quantity $n_{dv}$ of the PM protocol corresponds to the quantity
$n_{1v}$ of the ED protocol. For the same reason Eve cannot
distinguish the preparation of the signal states in the PM
protocol from that in the ED one. Furthermore, from the
definitions of $F_{conc}^{\beta}$ and $A_{fil}^{\beta}$, it can be
easily seen that the sequence of filters
$\{(1-F_v),A_{fil}^{\beta}\}$ is equivalent to the operator
$F_{conc}^{\beta}$ measured with a perfect detector. This also
implies that the quantity $n_{conc}$ of the PM protocol
corresponds to the quantity $n_{fil}$ of the ED protocol.

Regarding the security of the ED $\overline{\textrm{B92}}$, the
only point that deserves some care is the estimation of the
quantity $n_{1v}$ by means of the decoy states. In particular we
must show that this estimation is exponentially reliable. For that
we take inspiration from the estimation of the bit-error rate,
which follows closely the standard B92~\cite{TKI03,TL04}. The
number $n_{bit}$ of bit errors in the data pairs can be deduced
from the number $n_{err}$ of errors in the check pairs obtained in
the above ED-B92, Step (v). The argument is that in order to
obtain $n_{bit}$, Alice and Bob should perform $Z$ basis
measurements on the data pairs. Despite these measurements are not
really performed on data pairs, they are performed on the check
pairs. This is because the measurement $\mathcal{M}_{B92}^{\beta}$
is equivalent to a $Z$ measurement conditional on the outcome
$(1-F_v)$ of $Q_{v}$ and on the successful filtering
$A_{fil}^{\beta}$. Since in Step (ii) all the pairs are randomly
permuted, the check pairs can be seen as a classical random sample
of all the pairs remained after Step (iv). This leads to the
inequality $\left\vert n_{bit}-n_{err}\right\vert \leq
N\varepsilon$ which is exponentially reliable for large $N$.\\
We can apply the same argument to the decoy pairs. In order to
obtain $n_{1v}$ Alice and Bob should perform $X$ basis and QND
measurements on their data pairs. Although these measurements are
not really performed on data pairs, they are performed on decoy
pairs in Step (iv) of the ED protocol. Then, because of the random
permutation of Step (ii), the decoy pairs can be regarded as a
classical random sample of the $3N $ pairs prepared by Alice.
Hence we obtain that the estimation of $n_{1v}$ in our modified
B92 is exponentially reliable and can be used in the numerical
optimization of the phase-error upper bound. In the next Section
we will show the practical advantages of such a direct estimation.

\section{NUMERICAL SIMULATIONS}
\label{sec:numsim}

To see the practical advantages of the $\overline{\textrm{B92}}$
consider a channel with total loss rate $L$. If there is no
eavesdropping in the line it is natural to expect (see
Eq.\eqref{n1v}) that:
\begin{equation}\label{WCn1v2}
n_{1v} = \alpha ^{2}N L.
\end{equation}

\noindent Eq.\eqref{WCn1v2} can be experimentally verified in the
$\overline{\textrm{B92}}$; this represents the main advantage of
the new protocol. On the contrary, as already mentioned, Alice and
Bob can by no means verify Eq.\eqref{WCn1v2} in the standard B92,
so they must choose $n_{1v} = 0$, according to the worst-case
scenario described by Eq.\eqref{WCass}.

The settings of Eqs.\eqref{WCass},\eqref{WCn1v2} lead to the two
curves of Fig.1, which represent the gain $G$ as a function of the
loss rate $L$ for the traditional B92 and for the
$\overline{\textrm{B92}}$. The gain is given by $G=n_{fil}\left[
1-h\left(n_{bit}/n_{fil}\right) -h\left(
\overline{n}_{ph}/n_{fil}\right) \right] $, where
$\overline{n}_{ph}$ is the phase-error upper-bound, and $h$ is the
Shannon entropy~\cite{NC00}. The curves are drawn assuming, as
in~\cite{TL04}, a depolarizing channel with losses $\rho
\rightarrow L(1-p)\left\vert V\right\rangle \left\langle
V\right\vert +(1-L)\left[ (1-p)\rho +\sum\nolimits_{i=x,y,z}\left(
\sigma _{i}\rho \sigma _{i}\right)\right] $, where $\left\vert
V\right\rangle $ is the vacuum state, $\sigma _{i}$ are the Pauli
matrices and $p$, taken equal to 0.01 in our simulations, is the
depolarizing rate. The plot pertaining to the
$\overline{\textrm{B92}}$ is related to Eq.\eqref{WCn1v2}; on the
contrary that pertaining to the standard B92 comes from
Eq.\eqref{WCass}, and contains the same results given
in~\cite{TL04} for a B92 on a lossy channel. From the two plots is
apparent that our technique leads to a linear decrease of the gain
with the loss rate of the channel, rather than that, nearly
quadratic, of the standard B92. Only for $L=0$ the standard B92
features a higher gain. This is due to the presence of decoy
states that go discarded in the $\overline{\textrm{B92}}$.
%
%

\begin{figure}[tbp]
\includegraphics[width=8.5cm]{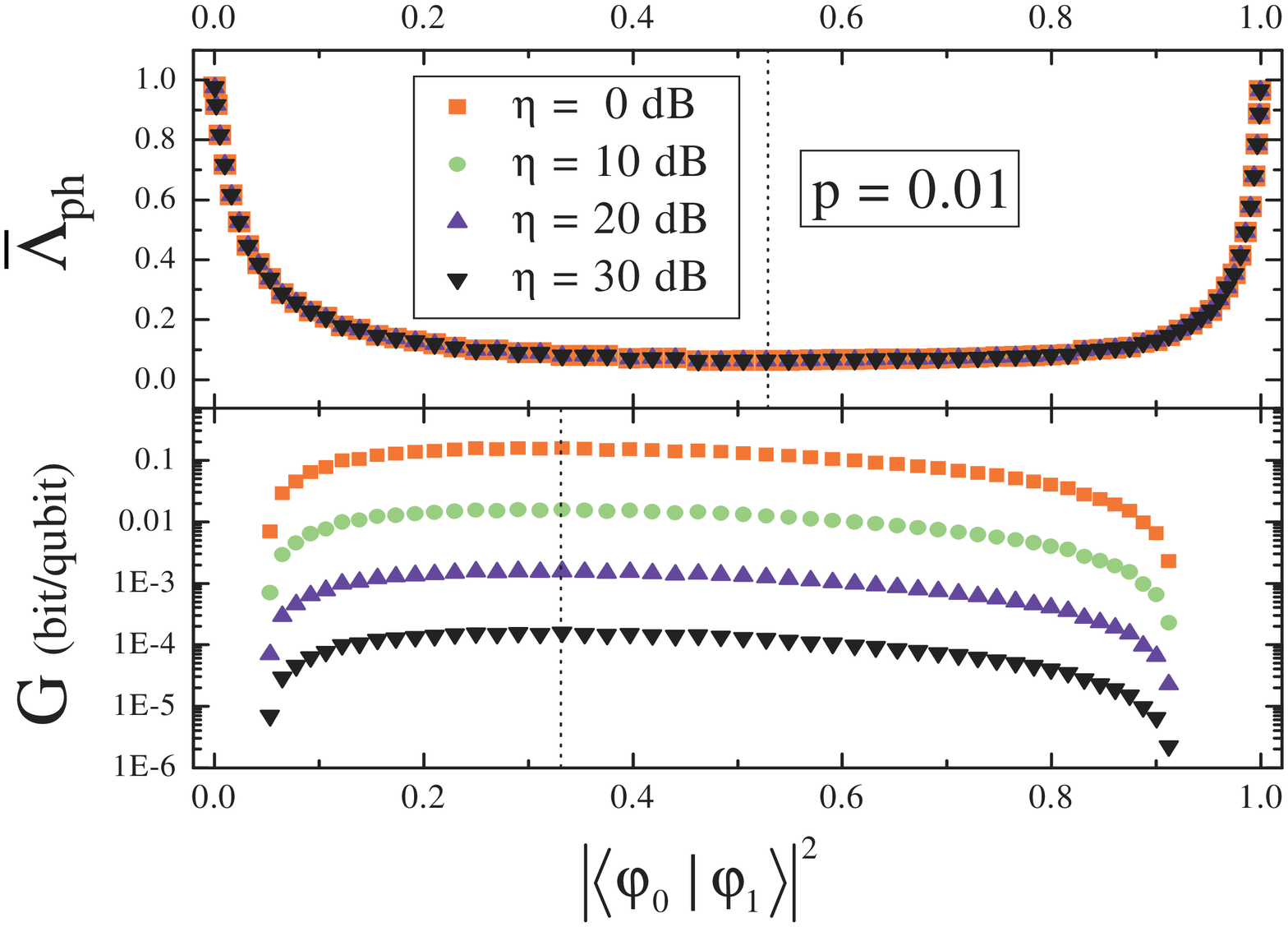}
\caption{(color online) Relative phase-error upper-bound
$\overline{\Lambda}_{ph}$ (top diagram) and secure gain $G$
(bottom diagram) of the $\overline{\textrm{B92}}$ versus the
squared scalar product of the two signal states $|\langle
\varphi_0 | \varphi_1\rangle|^2$, for channel transmission $\eta =
\{0, 10, 20, 30\}$ dB and depolarizing rate $p=0.01$. The vertical
dotted lines at 0.322 ($\theta=55.4^{\circ}$) and 0.538
($\theta=42.8^{\circ}$) mark respectively the maximum of $G$ and
the minimum of $\overline{\Lambda}_{ph}$.} \label{fig2}
\end{figure}

For the lossy depolarizing channel given above, the bit-error rate
can be easily found to be equal to $(1-L)p/3$, while that of
filtered states is $(1-L)[4p+3+(4p-3)\cos(2\theta)]/12$; hence
their ratio, i.e. the relative bit-error rate, is independent of
losses. In the $\overline{\textrm{B92}}$ even the upper-bounded
relative phase-error,
$\overline{\Lambda}_{ph}=\overline{n}_{ph}/n_{fil}$, is
independent of losses. This can be realized by observing the top
diagram of Fig.~\ref{fig2}, where we plotted
$\overline{\Lambda}_{ph}$ versus the squared scalar product of the
two signal states, $|\langle \varphi_0 | \varphi_1\rangle|^2$, for
several values of the total transmittance $\eta=1-L$. The points
pertaining to different values of the transmittance follow all the
same curve, thus demonstrating the independence of
$\overline{\Lambda}_{ph}$ from losses.\\
From the bottom diagram of Fig.~\ref{fig2}, containing the gain as
a function of $|\langle \varphi_0 | \varphi_1\rangle|^2$, we can
learn that, regardless of the transmittance, the maximum gain is
obtained when the two signal states are separated by an angle
$\theta \simeq 55.4^{\circ}$, while the minimum phase-error
upper-bound is obtained when $\theta \simeq 42.8^{\circ}$. This is
in sharp contrast with the standard theory of
B92~\cite{TL04},\cite{Tam08}, which assigns very small values to
$\theta$ to prevent the USD attack, thus reducing substantially
the final rate.

Since the relative bit-error and phase-error rates are
loss-independent, we can use the results of the lossless
B92~\cite{TKI03} to give an estimate of the single-photon
$\overline{\textrm{B92}}$ working distance when practical devices
are taken into account. It is found in~\cite{TKI03} that the
maximum depolarizing rate $p^{*}$ for which the gain of the
lossless single-photon B92 is still positive is $p^{\ast}=0.033$.
In real apparatuses, the depolarizing rate is given essentially by
detectors dark counts, which become dominant when the quantum
signal becomes too low. In~\cite{GYS04} the dark count probability
is $p_{dark}=1.7\times10^{-6}$, the attenuation of the fiber is
$\xi=0.21$ dB/Km, the mean detectors efficiency is $0.045$ and the
single-photon detection probability is $p_{s}(l)=0.045\times
10^{-\xi l/10}$, with $l$ the distance between the users. From the
inequality $p = p_{dark}/p_{s}(l)\leq p^{\ast}$ we can easily
obtain a working distance equal to about $140$ Km. This value can
be compared with the ones achieved by other protocols under
similar circumstances. For example it is known that the maximum
depolarizing rate of BB84 is $p^{*}=0.165$~\cite{SP00}, which
implies a working distance of about $173$ Km, while that of
SARG04~\cite{SARG} is $p^{*}=0.080$~\cite{TL06} with a
corresponding distance of $158$ Km. The difference between the
protocols depends crucially on their tolerance of the channel
noise, as exemplified by the above values of the depolarizing
rate. One way to make a protocol more tolerant to noise is to
improve its phase-error rate estimation. We accomplish this task
in the next section by slightly modifying the measuring apparatus
of the receiver Bob.

\section{IMPROVED PHASE-ERROR ESTIMATE}
\label{sec:phase-err}

With reference to the above-described ED protocol, in order to
directly measure $n_{ph} $, Alice and Bob should perform $X$ basis
measurements on their data pairs, and publicly compare their
results on the classical channel. However in this way their data
pairs could be no more used to distill a secret key, so this
procedure is usually substituted by the estimation of the
phase-error on a sub-sampling of the data pairs, the check pairs,
which are representative of the whole sample. This is what happens
in the BB84 for example~\cite{BB84,SP00}. Nevertheless such an
estimation procedure cannot be done in the standard B92 and
neither in our modified version $\overline{\textrm{B92}}$. The
reason is that although Alice prepares with a certain probability
the states in the $X$ basis, i.e. the decoy states, Bob never
measures them in the $X$ basis. By consequence, even in
$\overline{\textrm{B92}}$ the phase-error can not be directly
estimated and must be indirectly upper bounded from the values of
other quantities like $n_v$, $n_{err}$, $n_{fil}$ and $n_{1v}$,
through a numerical optimization algorithm.

As already mentioned, it is possible to further modify the B92 to
introduce such a direct estimation of the phase-error rate, and
make it more resistant to noise. The resulting protocol is the
$\overline{\overline{\textrm{B92}}}$. It consists in a random
switch of Bob's measurement between the $\mathcal{M}_{B92}^{\beta
} $ and a measurement in the $X$ basis. From the instances related
to the $X$ basis measurement the users can obtain a direct
estimation of the phase-error of the channel, as it happens in the
BB84. So, in particular, this is achieved by modifying the
following Steps of the given PM protocol:

\noindent (ii') Bob executes the measurement
$\mathcal{M}_{B92}^{\beta }$ \textit{with probability $1/2$ and
the measurement in the $X$ basis with probability $1/2$ and takes
note of outcomes;} [--].

\noindent (iii') After the quantum transmission, Bob tells Alice
the positions of vacuum, conclusive and inconclusive counts
\textit{and those of his measurements in the X basis}; [--]. She
announces the estimated quantities to Bob together with the
positions \textit{and the values} of decoy bits. \textit{All the
instances in which Alice prepared signal states and Bob measured
in the X basis are discarded}.

\noindent (iv') \textit{Bob estimates the number of errors
$n_{ph}$ in the decoy instances in which Alice prepared
$|1_x\rangle$ $(|0_x\rangle)$ and Bob detected $|0_x\rangle$
$(|1_x\rangle)$}; [--].

\noindent (vi') From $n_{err}$ the users estimate the number of
bit errors $n_{bit}$.

The slight increase in the complexity of Bob's measurement, and
the decrease in the final rate entailed by the new Step (iii'),
are compensated by the benefits of a better tolerance to the
channel noise. By assuming again a depolarizing channel, it can be
easily seen that the tolerable depolarizing rate $p^{*}$ depends
directly on the angle $\theta$ between the two signal states: the
greater $\theta$ the larger $p^{*}$. This is summarized in
Table~\ref{tab1}.
\medskip
\begin{table}[h!]
  \centering
  \begin{tabular}{|c|c|c|c|c|c|c|c|c|c|}
  \hline
  $\theta$ & $10^{\circ}$ & $20^{\circ}$ & $30^{\circ}$ & $40^{\circ}$ & $50^{\circ}$ & $60^{\circ}$ & $70^{\circ}$ & $80^{\circ}$ & $90^{\circ}$\\
  \hline
  $p_{(\%)}^{\ast}$ & $0.4$ & $1.5$ & $3.4$ & $5.9$ & $8.6$ & $11.5$ & $14.0$ & $15.8$ & $16.5$ \\
  \hline
  \end{tabular}
  \caption{The maximum depolarizing rate tolerable by the $\overline{\overline{\textrm{B92}}}$,
  and the corresponding angle $\theta$ between the two signal states. The depolarizing rates for
  BB84~\cite{BB84} and SARG04~\cite{SARG} are respectively $16.5\%$~\cite{SP00} and $8.04\%$~\cite{TL06}. } \label{tab1}
\end{table}
\medskip
\noindent It can be seen that the tolerable depolarizing rate can
be increased up to the BB84 level, and well above the SARG04
threshold. However we note incidentally that a similar solution
for a direct phase-error estimation can be applied to the SARG04
protocol by modifying only its classical data processing. Let us
also notice that the unconditional security of the
$\overline{\overline{\textrm{B92}}}$ follows closely that of the
$\overline{\textrm{B92}}$; in fact the decoy instances are used as
a classical sampling to provide an exponentially reliable bound to
the phase errors.

\section{CONCLUSION}
\label{sec:conclusion}

In conclusion we have shown that a pair of uninformative states
can be introduced in the single-photon B92 protocol in order to
remove its high dependance on the losses and noise of a
communication channel. In particular, without modifying the
receiver's box, the technique can prevent at the root an USD
attack by Eve. Furthermore, with a slight modification of Bob's
measurement, the technique allows a direct estimation of the
number of phase errors, thus increasing the robustness of the
protocol against external sources of noise.\\
The results are of theoretical interest since they solve at the
root the long-standing problem of the unambiguous state
discrimination of the two single-photon B92 signal states. In
fact, other protocols based on similar principles could benefit of
our analysis~\cite{DB04,BD05}. Furthermore, although the results
are limited to the single-photon case, they are easily exportable
to a realistic
scenario by applying the well established decoy-state technique~\cite{Hwa03,LMC05,Wan05}.\\
The presented method can be extended to more than two
uninformative states, with the potential of diverting other
USD-based attacks~\cite{Koa05,LM05,LCD+07}, possibly related to a
non-ideal equipment of the users. Another option is to adapt the
proposed solution to the B92 with a not-so-strong reference
pulse~\cite{Tam08}; this would allow to use in that protocol a
wider angle between the signal states, thus reducing the problem
of a precise phase stabilization.

\section*{ACKNOWLEDGMENTS}

We thank Paolo~Tombesi, Norbert~L\"{u}tkenhaus and Rupesh Kumar
for stimulating discussion. One of us (M.L.) is grateful to
Hoi-Kwong Lo for his hospitality in the Quantum Information and
Quantum Control group, Department of Electrical \& Computer
Engineering, University of Toronto, Canada.


\end{document}